\documentclass[lettersize,journal]{IEEEtran}
\usepackage{amsmath,amsfonts}
\usepackage{algorithmic}
\usepackage{algorithm}
\usepackage{array}
\usepackage[caption=false,font=normalsize,labelfont=sf,textfont=sf]{subfig}
\usepackage{textcomp}
\usepackage{stfloats}
\usepackage{url}
\usepackage{verbatim}
\usepackage{graphicx}
\usepackage{cite}
\usepackage{caption}

\newcommand{\cmt}[1]{}

\usepackage{makecell}
\usepackage[normalem]{ulem}
\usepackage{caption}
\usepackage{enumitem}
\usepackage{float}
\usepackage{amsmath}
\usepackage{booktabs}
\usepackage{xcolor}
\usepackage{graphicx}
\usepackage{cuted}

\newcommand{\reffig}[1]{{Fig.~\ref{fig:#1}}}
\newcommand{\reftab}[1]{{Tab.~\ref{tab:#1}}}
\newcommand{\refsec}[1]{{Sec.~\ref{sec:#1}}}
\newcommand{\refeqn}[1]{{Eqn.~(\ref{eqn:#1})}}

\begin{document}

\title{X2Video: Adapting Diffusion Models for Multimodal Controllable Neural Video Rendering}

\author{Zhitong Huang,
\thanks{- Zhitong Huang is with City University of Hong Kong, Hong Kong, China. E-mail: luckyhzt@gmail.com}
\and
Mohan Zhang,
\thanks{- Mohan Zhang is with WeChat, Tencent Inc, Shenzhen, China. E-mail: zeromhzhang@tencent.com}
\and
Renhan Wang,
\thanks{- Renhan Wang is with Manycore Tech Inc., Hangzhou, China. E-mail: xichen@qunhemail.com}
\and
Rui Tang,
\thanks{- Rui Tang is with Manycore Tech Inc., Hangzhou, China. E-mail: ati@qunhemail.com}
\and
Hao Zhu,
\thanks{- Hao Zhu is with Manycore Tech Inc., Hangzhou, China. E-mail: hunyuan@qunhemail.com}
\and
Jing Liao$^{*}$
\thanks{- Jing Liao is with City University of Hong Kong, Hong Kong, China. E-mail: jingliao@cityu.edu.hk}
\thanks{*: Corresponding author}
}

\markboth{}%
{}
\IEEEpubid{}

\maketitle

\begin{strip}
    \vspace{-115pt}
    \centering
    \includegraphics[width=\textwidth]{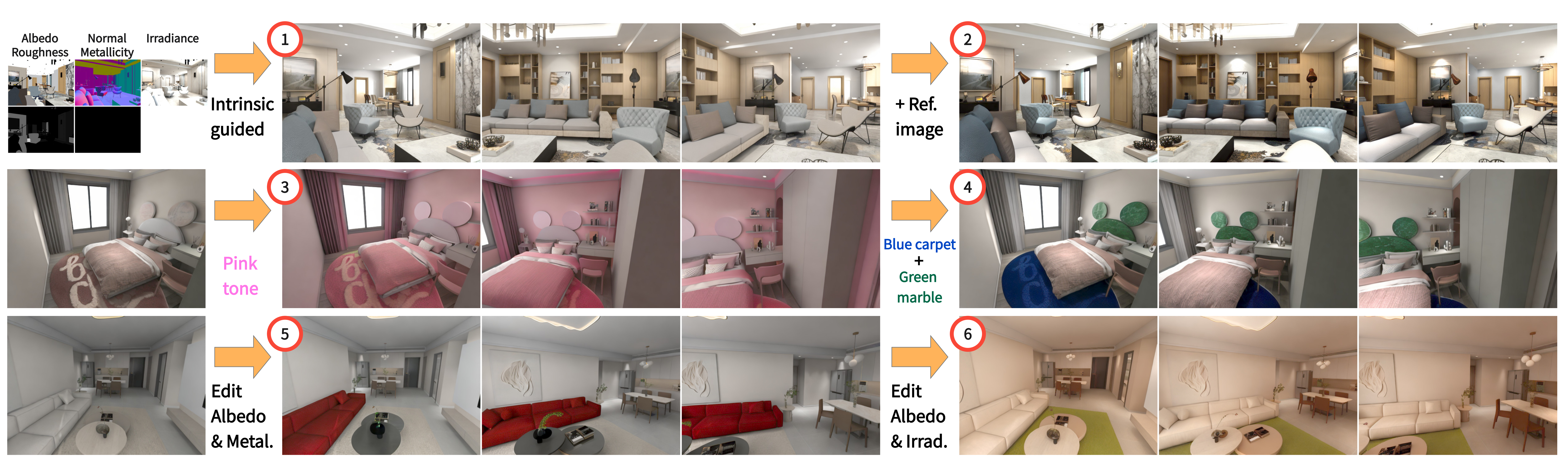}
    \captionof{figure}{Our X2Video can generate long, temporally consistent, and photorealistic videos with multi-modal controls including intrinsic channels ($1^\text{st}$ example), reference image ($2^\text{nd}$ example), global text control ($3^\text{rd}$ example), and text control on local regions ($4^\text{th}$ example). Additionally, X2Video also supports editing color, material, texture or lighting through parametric tuning, as in the $4^\text{th}$ and $5^\text{th}$ example.}
    \label{fig:teaser}
\end{strip}

\begin{abstract}
We present X2Video, the first diffusion model for rendering photorealistic videos guided by intrinsic channels including albedo, normal, roughness, metallicity, and irradiance, while supporting intuitive multi-modal controls with reference images and text prompts for both global and local regions.
The intrinsic guidance allows accurate manipulation of color, material, geometry, and lighting, while reference images and text prompts provide intuitive adjustments in the absence of intrinsic information.
To enable these functionalities, we extend the intrinsic-guided image generation model XRGB to video generation by employing a novel and efficient \textit{Hybrid Self-Attention}, which ensures temporal consistency across video frames and also enhances fidelity to reference images.
We further develop a \textit{Masked Cross-Attention} to disentangle global and local text prompts, applying them effectively onto respective local and global regions.
For generating long videos, our novel \textit{Recursive Sampling} method incorporates progressive frame sampling, combining keyframe prediction and frame interpolation to maintain long-range temporal consistency while preventing error accumulation.
To support the training of X2Video, we assembled a video dataset named \textit{InteriorVideo}, featuring 1,154 rooms from 295 interior scenes, complete with reliable ground-truth intrinsic channel sequences and smooth camera trajectories.
Both qualitative and quantitative evaluations demonstrate that X2Video can produce long, temporally consistent, and photorealistic videos guided by intrinsic conditions.
Additionally, X2Video effectively accommodates multi-modal controls with reference images, global and local text prompts, and simultaneously supports editing on color, material, geometry, and lighting through parametric tuning.
Upon acceptance, we will publicly release our model and dataset. Project page available at: \textit{\textcolor{blue}{ \url{https://luckyhzt.github.io/x2video}}}.
\end{abstract}


\section{Introduction}
To synthesize photorealistic images and videos, the traditional Graphics Pipeline relies on physically based rendering (PBR) \cite{pharr2023physically}, which incorporates accurate simulation of light interactions with surfaces and materials, meticulously accounting for properties like diffuse reflection, specular highlights, and subsurface scattering to achieve highly realistic visual results. Furthermore, PBR supports precise adjustment of rendering effects through professional and parametric tuning of geometry, materials, and lighting. However, this process demands expert skills in 3D software and lacks flexibility and intuitive interaction, making it particularly difficult for non-professional users to achieve desired results.
Additionally, PBR methods are computationally intensive, which poses challenges such as increased processing time and resource demands, particularly for high-resolution and high-quality image sequences and video rendering.

Generative models, particularly large diffusion models \cite{ldm,svd}, have emerged as a novel approach to producing realistic images and videos, differing significantly from traditional rendering methods by learning to generate content from data rather than simulating physical processes.
Recent works \cite{uni-render, xrgb, diffusion-render} have introduced intrinsic guidance to diffusion models to generate photorealistic content with accurate lighting, material, and geometry. This intrinsic guidance allows the models to understand and manipulate the underlying physical properties of a scene, moving beyond mere physical simulation to a more informed synthesis.
While Uni-Render \cite{uni-render} and Diffusion-Render \cite{diffusion-render} generate rendered objects and videos with complete intrinsic information, XRGB \cite{xrgb}  appears to be the pioneer in introducing intuitive textual control to AI rendering.
Unlike PBR, which necessitates a full and complete set of scene specifications, XRGB can generate images from partial intrinsic information, granting the model the freedom to hallucinate a plausible result according to intuitive textual controls.
This enables intuitive controls on colors, materials, and lighting based on textual prompts, providing a more accessible alternative to professional parametric editing.
Despite these advantages, textual control in XRGB is limited only to the global region when an entire intrinsic channel is missing, which cannot suit applications requiring textual controls over local regions and objects. 
Additionally, XRGB does not support using reference images as control when text descriptions fall short of precisely conveying the intended colors, textures, or materials.

To address these limitations and enhance the flexibility and precision of AI rendering, we introduce X2Video, the first intrinsic-guided video rendering framework with intuitive multi-modal controls. This video diffusion model leverages intrinsic information, albedo (base color), normal (surface orientation), irradiance (ambient lighting), roughness, and metallicity as input to guide the video generation process, while supporting multi-modal controls through reference images and text prompts applicable to both local and global regions. 
Our framework enables text-based controls to compensate for missing intrinsic information. For instance, a global text prompt can regulate lighting when the irradiance channel is absent ($3^\text{rd}$ example in \reffig{teaser}), while textual control on color and material of local regions is enabled with local masks on albedo, metallicity, and roughness ($4^\text{th}$ example in \reffig{teaser}).
In cases where text prompts are insufficient to accurately describe the desired color and material details, our framework provides an alternative to use the first frame as a reference to indicate the details of the local regions and global rendering styles ($2^\text{nd}$ example in \reffig{teaser}).
These comprehensive multi-modal controls ensure precise, intuitive, and highly flexible video rendering across diverse scenarios.

The proposed functionalities require that X2Video must: 1) precisely capture intrinsic conditions, 2) ensure strong temporal consistency across rendered video frames, 3) faithfully utilize reference images, 4) disentangle and apply the text prompts effectively to both global and local regions, and 5) generate long, coherent videos.
To achieve these, we initially leverage the pretrained XRGB model as our base model to inherit its knowledge to translate intrinsic data into photorealistic images.
This foundation provides basic capabilities to render individual frames.
Building upon this, we introduce a novel \textit{Hybrid Self-Attention} mechanism, integrating Reference Attention \cite{lvcd} to enhance fidelity to reference images and Multi-Head Full (MHF) Temporal Attention to thoroughly understand inter-frame relationships, ensuring temporal consistency.
To handle global and local text prompts, we implement a novel \textit{Masked Cross-Attention} to apply prompts separately to their respective global and local regions.
Considering the extension to long video generation, we propose a novel \textit{Recursive Sampling} scheme to progressively render frames by sampling from long temporal intervals and recursively interpolating intermediate frames. This mechanism successfully eliminates the error accumulation problem in autoregressive sequential sampling and enhances the long-range temporal consistency.
Finally, to support the training of our proposed video model, we create a new video dataset, \textit{InteriorVideo}, containing 1154 rooms across 297 different interior scenes with rendered video frames and ground-truth sequences of intrinsic channels.

In summary, our key contributions are as follows:
\begin{itemize}[leftmargin=11pt, topsep=0pt]
    \item We introduce X2Video, the first intrinsic-guided video diffusion model to render photorealistic videos with intuitive multi-model controls, which leverages intrinsic channels, reference images, and text prompts to intuitively control color, material, lighting, and geometry in both local and global regions.
    \item We propose a novel Hybrid Self-Attention layer that combines Multi-Head Full (MHF) Temporal Attention for improved temporal consistency and Reference Attention for enhanced fidelity to image reference, as well as a novel Masked Cross-Attention layer to effectively handle both global and local text controls.
    \item We present a novel Recursive Sampling scheme to produce long, temporally consistent videos.
    \item We create a video dataset of interior scenes, consisting of rendered video frames and ground-truth intrinsic channel sequences, with complete intrinsic channels and smooth camera trajectories.
\end{itemize}

\section{Related Works}

\subsection{Intrinsic-Guided Diffusion Models}
Generative models, particularly diffusion models, have garnered significant attention due to their remarkable ability to synthesize high-quality, photorealistic images from text prompts \cite{ldm, dalle, imagen}. 
Despite their success, these models often lack fine-grained control over structural composition, material properties, and lighting conditions. 
To enhance controllability, frameworks such as ControlNet \cite{controlnet} and T2I-Adapter \cite{t2iadapter} introduce conditional guidance like edge maps, depth, or segmentation masks. 
However, they remain limited in precisely manipulating physical material attributes and complex lighting effects.

To address this, previous works \cite{uni-render, xrgb, diffusion-render} have introduced intrinsic channels, including albedo, normal, irradiance, roughness, and metallicity, to diffusion models for image and video generation.
This enables the generation of high-quality content with physically accurate materials and lighting.
For instance, Uni-Render \cite{uni-render} proposes a dual-stream diffusion framework that jointly supports forward rendering (image synthesis) and inverse rendering (decomposition of real images into intrinsic components) for single objects under controlled settings. 
Diffusion-Render \cite{diffusion-render}, on the other hand, extends this idea to video sequences, introducing a video diffusion framework capable of both forward and inverse rendering for synthetic and real-world videos.
Among these approaches, XRGB \cite{xrgb} stands out as the first AI-rendering framework to support intuitive textual control over global scene properties, effectively combining the precision of conventional PBR with the flexibility of generative models. 
Nevertheless, XRGB only allows global textual controls and does not support localized text prompts to modify specific local regions or objects. 
Moreover, it lacks support for reference images as an additional intuitive control, which can be crucial when textual descriptions are insufficient to describe the desired details or styles.
To overcome these shortcomings, we propose a multi-modal extension to intrinsic-guided models that supports reference image and text prompts on both local and global regions, enabling more precise and flexible rendering.

\subsection{Video Diffusion Models}
With the advancement of video diffusion models, a variety of video synthesis frameworks have emerged for tasks such as text-to-video generation \cite{cogvideox, gridvideo}, image-to-video \cite{svd, animatediff, i2vadapter}, and video generation conditioned on depth maps \cite{controlvideo} or linearts \cite{lvcd}. 
Despite these advances, existing methods are not designed to generate videos from sequences of intrinsic image components and thus struggle to preserve accurate and realistic material and lighting.
To address this gap, we propose the first video diffusion model designed for synthesizing videos from sequences of intrinsic channels, leveraging the pretrained image diffusion model, XRGB, which already incorporates intrinsic channel controls.

Most prior efforts \cite{svd, animatediff} to adapt pretrained image diffusion models for video generation introduce additional temporal layers, such as 3D convolutions or temporal attention layers.
However, these layers differ significantly from the original image models and typically require training on large-scale video datasets from scratch, limiting their feasibility in tasks with insufficient data.
Furthermore, the typical combination of 3D convolution and one-dimensional temporal attention restricts receptive fields, failing to ensure full attention for long-range consistency in video processing.
To overcome these limitations, we introduce a novel Hybrid Self-Attention that enables full temporal attention across generated frames, where all additional layers are initialized from the original spatial self-attention layers, enhancing the efficiency of the extension to video models.

\subsection{Text Control in Diffusion Models}
Text prompts have become a fundamental modality for intuitive control in generative models, widely adopted in both image \cite{ldm, imagen} and video synthesis \cite{cogvideox}. 
Typically, these prompts describe the global semantic content of the target output but provide limited means to specify the semantics and content of particular local regions. 
To enable localized control, prior works have explored manipulating the cross-attention layers within diffusion models to associate specific words in the prompt with corresponding local regions.
For instance, Zhang et al. \cite{continuous_layout} and Hertz et al. \cite{prompt2prompt} propose methods that reweight the cross-attention matrix during inference, effectively aligning certain text tokens with corresponding local regions to enable image editing.
Unlike these inference-only methods, GLIGEN \cite{gligen} introduces a gated cross-attention that explicitly conditions generation on spatial grounding by incorporating bounding boxes during training, which ensures reliable object localization from layout inputs. 
Similarly, LayoutDiffusion \cite{layoutdiffusion} enhances spatial control by injecting layout information directly into the diffusion process through dual cross-attention blocks that jointly process textual prompts and spatial layouts. This design ensures that generated objects not only match the described semantics but also adhere to the specified spatial layout.

\section{Datasets}

\subsection{Datasets with intrinsic channels}
\label{sec:dataset}
Open-source datasets \cite{HyperSim, InteriorVerse, InteriorNet} provide images of interior scenes accompanied by corresponding intrinsic channels. 
However, as summarized in \reftab{dataset}, these datasets exhibit notable limitations, such as inaccuracies in the provided intrinsic channels and discontinuous camera trajectories that lack smoothness, which hinders their effectiveness for video-based tasks.
To address these shortcomings, we introduce \textit{InteriorVideo}, the first video dataset featuring smooth, natural camera trajectories and a complete set of reliable ground-truth intrinsic channels.

\begingroup
\setlength{\tabcolsep}{4pt}
\begin{table}[t]
    \footnotesize
    \renewcommand{\arraystretch}{1.6}
    \centering
    \begin{tabular}{ccccccc}
         \toprule
         &Albedo & Normal &  Roughness &  Metallic. & Irrad. & \makecell{Smooth \\ Cameras} \\ \hline 
         HyperSim&\textcolor{orange}{\checkmark}&  \textcolor{green}{\checkmark}&  \textcolor{red}{$\boldsymbol{\times}$}&   \textcolor{red}{$\boldsymbol{\times}$}& \textcolor{green}{\checkmark}& \textcolor{red}{$\boldsymbol{\times}$}\\
         InteriorVerse&\textcolor{green}{\checkmark}&  \textcolor{green}{\checkmark}&  \textcolor{orange}{\checkmark}&   \textcolor{orange}{\checkmark}& \textcolor{red}{$\boldsymbol{\times}$}& \textcolor{red}{$\boldsymbol{\times}$}\\
         InteriorNet&\textcolor{green}{\checkmark}&  \textcolor{green}{\checkmark}&  \textcolor{red}{$\boldsymbol{\times}$}&   \textcolor{red}{$\boldsymbol{\times}$}& \textcolor{orange}{\checkmark}& \textcolor{orange}{\checkmark}\\
         \makecell{InteriorVideo \\ (Ours)}&\textcolor{green}{\checkmark}&  \textcolor{green}{\checkmark}&  \textcolor{green}{\checkmark}&   \textcolor{green}{\checkmark}& \textcolor{green}{\checkmark}& \textcolor{green}{\checkmark}\\ 
         \toprule
    \end{tabular}
    \caption{Comparisons between interior datasets. Each channel is marked as "available" (\textcolor{green}{\checkmark}), "unavailable" (\textcolor{red}{$\boldsymbol{\times}$}), and "available but unreliable" (\textcolor{orange}{\checkmark})}
    \label{tab:dataset}
\end{table}
\endgroup

To construct our dataset, we first decompose each interior scene into individual rooms. Within each room, we initialize the camera at a height of 1.7 meters (approximating human eye level) at randomly selected positions located at the corners or midpoints of room boundaries. The camera is oriented toward the geometric center of the room to ensure meaningful scene coverage.
At each step along the trajectory, we generate 900 candidate camera positions within a local neighborhood, excluding any positions that result in collisions. For each valid candidate, we compute the number of visible objects $N_i$ within its field of view, and sample the next camera position with weights defined as $w_i = N_i^2$. This encourages the camera to move toward regions with richer object content and higher visual complexity.
The camera traverses the scene at an average speed of 0.05 m/s and yields video sequences at approximately 10 frames per second (FPS), resulting in an effective speed of 0.5 m/s.
Once the camera trajectories are established, we render the RGB frames along with complete ground-truth intrinsic channels (albedo, normal, roughness, metallicity, and irradiance).
To enhance the quality of the irradiance channel, we apply denoising with the Open Image Denoise library (OIDN) \cite{oidn}. 
The rendered video sequences exhibit motion dynamics with an average displacement of 3.16 pixels between consecutive frames, as measured by pixel-wise optical flow.
\begin{figure*}[t]
    \centering
    \includegraphics[width=1.0\linewidth]{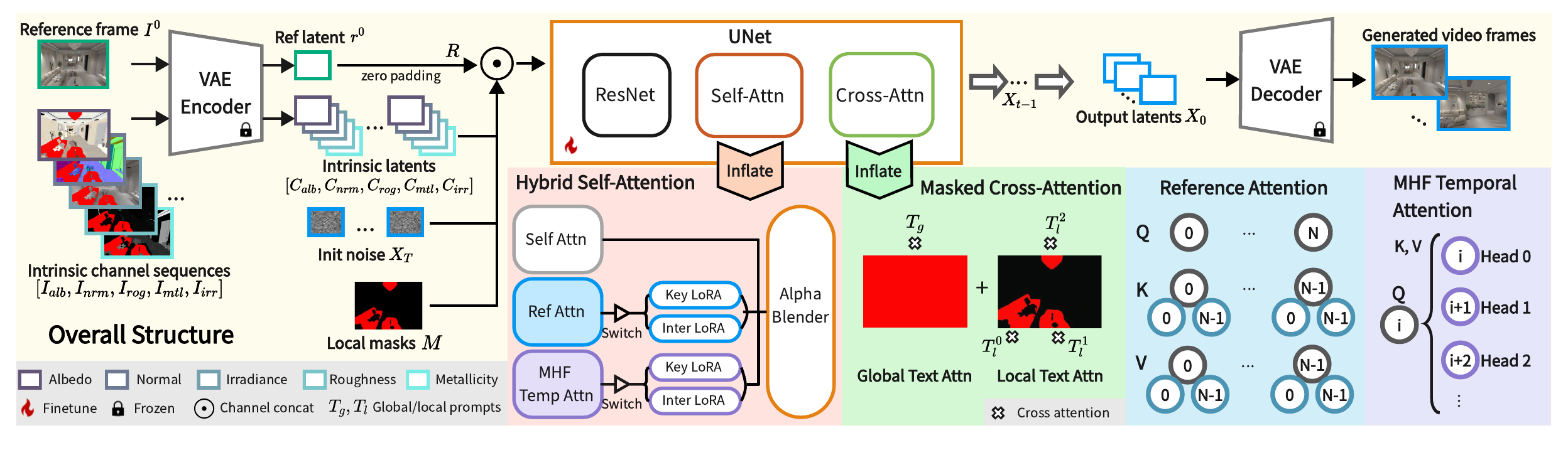}
    \caption{Overall structure of our framework. Given a sequence of intrinsic channels (with optional multi-modal conditions including a reference frame, global text, and local masked text), our model can generate temporally consistent video. Hybrid Self-Attention is proposed to enhance temporal consistency and fidelity to the reference. Masked Cross-Attention is proposed to effectively apply global and local text prompts on corresponding global and local regions.}
    \label{fig:overall_structure}
\end{figure*}

\section{Methodology}

\subsection{Preliminary: Diffusion Models}
We employ a pretrained XRGB model \cite{xrgb}, which is built upon the text-to-image latent diffusion model Stable Diffusion (SD) 2.1 \cite{ldm}. 
Our framework follows the same training objective as SD 2.1 with v-prediction \cite{v-pred}, which parameterizes the model to predict a hybrid of the noise and original signal. 
The training loss is defined as:
\begin{equation}
\label{eqn:loss}
\begin{aligned}
    L_\theta &= \| \text{v}_t - \hat{\text{v}}_\theta  \left( t, x_t, T_g, T_l, r^0, C \right)  \|^2_2  \\
    \text{v}_t &= \sqrt{\bar{\alpha}_t} \epsilon - \sqrt{1 - \bar{\alpha}_t} x_0
\end{aligned}
\end{equation}
where $\hat{\text{v}}_\theta$ denotes the predicted output by the diffusion model with parameters $\theta$. 
The model takes as input the diffusion timestep $t$, the noised latent representation $x_t$, global and local text prompts $T_g$ and $T_l$, the reference frame $r^0$, and the intrinsic channels $C$. 
The target $\text{v}_t$ is computed from the added normal noise $\epsilon$ and the original (clean) latent $x_0$, weighted by the cumulative noise schedule parameters $\bar{\alpha}_t$.

\subsection{Model Overview}
As illustrated in \reffig{overall_structure}, our goal is to generate a temporally coherent video sequence $\hat{I}^{[0:N)}$ conditioned on a set of intrinsic channel sequences $[ I_c^{[0:N)} ]$, where each channel is defined as $[I_c] = \left[ I_{alb}, I_{nrm}, I_{irr}, I_{rog}, I_{mtl} \right]$ (representing albedo, normal, irradiance, roughness, and metallicity, respectively).
Additionally, we incorporate optional multi-modal conditional inputs: a reference image $I_0$, a global text prompt $T_g$, and a set of multiple local text prompts $T_l^{[0:P)}$ along with corresponding spatial masks $M^{[0:P)}$.
First, we encode the reference frame $I_0$ and the intrinsic frames $I_c^{[0:N)}$ using a frozen VAE encoder.
This process yields the reference latent representation $r^0$ and the intrinsic latents $\left[ C_{alb}, C_{nrm}, C_{irr}, C_{rog}, C_{mtl} \right]$.
To align the reference latent with the temporal dimension of the intrinsic latents, we pad $r^0$ with zeros along the temporal axis to form the reference latent array $R$.
Next, we construct the input to the diffusion model by concatenating the initial noisy latent $x_T$, the intrinsic latents $C$, the padded reference latents $R$, and the local masks $M^{[0:P)}$ along the channel dimension. 
This concatenated tensor serves as input to the UNet backbone of the diffusion model for iterative denoising.
Within the UNet's attention blocks, we employ Masked Cross-Attention to condition the generation process on both global and local text prompts. 
Specifically, the global text prompt $T_g$ is attended to all spatial locations, while each local text prompt $T_l^p$ is only accessible within the spatial region defined by its corresponding mask $M^p$, enabling localized semantic control.
The output of the UNet, representing the predicted clean latent $X_0^{[0:N)}$, is subsequently decoded through a frozen VAE decoder to reconstruct the final video sequence $\hat{I}^{[0:N)}$ in pixel space.

In the following sections, we first describe the architecture of Hybrid Self-Attention layers that extend the pre-trained 2D self-attention layers to handle spatio-temporal modeling and enhance fidelity to the input reference image (\refsec{hybrid_attn}). 
We then detail the implementation of Masked Cross-Attention for both global and local textual control (\refsec{mask_attn}). 
Finally, we introduce Recursive Sampling for generating longer video sequences (\refsec{recursive_sample}).

\subsection{Hybrid Self-Attention}
\label{sec:hybrid_attn}
As illustrated in \reffig{overall_structure}, we enhance the conditional strength of the reference frame and improve the temporal consistency of our model by transforming the original self-attention layers into a new structure called Hybrid Self-Attention. 
In this approach, the outputs from the standard self-attention mechanism are augmented with contributions from two additional components: Reference Attention \cite{lvcd} and Multi-Head Full (MHF) Temporal Attention. 
Specifically, the Reference Attention focuses on aligning features derived from the reference frame $r^0$, thereby strengthening the influence of the reference image across the generated sequence. Meanwhile, MHF Temporal Attention captures long-range dependencies across frames, ensuring coherent temporal dynamics throughout the video.
These components are integrated using the Alpha Blender module to produce the final output of the Hybrid Self-Attention layer.

\textbf{Reference Attention.}
As evidenced in \cite{lvcd}, Reference Attention enhances long-range correspondence with the input reference frame, improving temporal consistency and preserving higher fidelity to the visual content in the reference image. 
This mechanism enables each generated frame to maintain strong structural and appearance alignment with the reference.

We implement Reference Attention to enable full attention interaction between the reference frame features and those of each subsequent generated frame. Formally, it is defined as:
\begin{equation}
\begin{aligned}
    & \text{RefAttn} \left( f^i \right) = \text{Attn} \left(  w_r^q(f^i), w_r^k(f^0), w_r^v(f^0)  \right) \\
\end{aligned}
\end{equation}
where $\text{Attn}(\cdot)$ denotes the standard scaled dot-product multi-head attention mechanism \cite{attention}, $f^i$ represents the intermediate feature map at frame $i$ input to the Hybrid Self-Attention layer, and $f^0$ denotes the corresponding feature map from the reference frame. 
The query, key, and value projection weights ($w_r^q$, $w_r^k$, and $w_r^v$) are initialized from the pre-trained weights of the original self-attention layers, ensuring stable training and effective transfer of learned knowledge. 
By using the reference frame as the sole source for keys and values, the model attends to consistent reference content to enhance fidelity to the input reference frame and the long-range temporal consistency to the first frame.

\textbf{Multi-Head Full Temporal Attention.}
While reference attention improves long-range consistency with the first reference frame, there is still a need for temporal attention to capture correlations among all generated frames.
A common approach is to integrate a 1-dimensional temporal attention layer after the self- and cross-attention layers.
However, this temporal attention has a fundamentally different structure from the original attention layers, requiring training from scratch.
This can be problematic, as it often necessitates substantial amounts of video data, which may not be available for tasks with limited data.

Therefore, we propose Multi-Head Full (MHF) Temporal Attention, which utilizes a multi-head mechanism to achieve full temporal attention with low computational and training costs.
We first initialize the weights for the query, key, and value mappings in MHF Temporal Attention to the original self-attention weights and encode the input features into multiple heads:
\begin{equation}
    Q^{i,[0:H)}_t, \, K^{i,[0:H)}_t, \, V^{i,[0:H)}_t = w_t^q(f^i), \, w_t^k(f^i), \, w_t^v(f^i)
\end{equation}
where $\{Q,K,V\}^{i,[0:H)}_t$ denotes the query, key, and value matrices for frame $i$ with $H$ heads. Then we modify the standard multi-head self-attention from:
\begin{equation}
    \text{Attn} \left(  Q^{i, h}_t, K^{i, h}_t, V^{i, h}_t  \right) \;\;\; \text{for} \; h \in [0, H)
\end{equation}
to our MHF Temporal Attention as follows:
\begin{equation}
\begin{aligned}
    \text{TempAttn} \left( f^i \right) &= \text{Attn} \left( Q^{i,h}_t, K^{i_t,h}_t, V^{i_t,h}_t \right) \\ 
    \text{for} & \; h \in [0, H) \;\;\; i_t = (i + h) \mod{N}
\end{aligned}
\end{equation}
where $N$ represents the total number of frames, and $i_t$ is the frame index that interacts with frame $i$ through the $h^{\text{th}}$ head. 
This design enables each head of $Q_t^i$ to interact with the keys and values of different frames, achieving full attention across all input frames.
As shown in the purple region of \reffig{overall_structure}, the $0^{\text{th}}$ head calculates self-attention, while the $1^{\text{st}}$ head computes full temporal attention with the next frame (i.e., frame $i+1$), and so forth.
Consequently, MHF Temporal Attention maintains the original self-attention mechanism's computational complexity while enabling comprehensive temporal attention across all input frames.

\textbf{Alpha Blender.}
Finally, we combine the outputs from the self-attention, reference attention, and MHF Temporal Attention using the Alpha Blender module:
\begin{equation}
\begin{aligned}
    \text{HybridAttn}(f^i) & = (1 - \alpha_r - \alpha_t) \cdot \text{SelfAttn} (f^i) \\ 
                        & + \alpha_r \cdot \text{RefAttn} (f^i) + \alpha_t \cdot \text{TempAttn} (f^i)
\end{aligned}
\end{equation}
where $\alpha_r$ and $\alpha_t$ are learnable scalar coefficients that control the contributions of the reference attention and temporal attention, respectively. 
These weights are shared across spatial positions and attention heads within each layer but are allowed to vary across different layers of the network, enabling layer-specific adaptation of attention routing.

To ensure stable training and maintain the pre-trained knowledge of the original 2D diffusion model at initialization, we initialize both $\alpha_r$ and $\alpha_t$ to zeros. 
This guarantees that, at the start of training, the Hybrid Self-Attention layers behave identically to the original self-attention layers, producing the same outputs as the pretrained image model. 
The model then gradually learns to activate the reference and temporal attention as training progresses.

\subsection{Masked Cross-Attention}
\label{sec:mask_attn}
To effectively disentangle the global text prompt and the set of local text prompts and apply them to their corresponding global and local regions specified by masks, we design a novel Masked Cross-Attention.
As illustrated in the green region of \reffig{overall_structure}, global text attention is computed as the conventional cross-attention between the entire image region and the global text prompt $T_g$. 
In contrast, local text attention is computed as the cross-attention between each local text prompt $T_l^p$ and its corresponding masked region $M^p$ within the frame (highlighted in red), with other unmasked regions set to 0.
The final output of the Masked Cross-Attention is computed as the masked summation of the global and local text attentions:
\begin{equation}
\label{eqn:mask_crossattn}
    \text{MCAttn}(f_i) = \text{CAttn}(f_i,T_g) + \sum_p  M^p \cdot \text{CAttn} \left( f_i,T_l^p \right)
\end{equation}
where $\text{CAttn}(\cdot)$ denotes the standard cross-attention between image and text features, and the mask $M^p$ assigns a value of 1 to pixels within the region associated with text prompt $T_l^p$. 
This formulation ensures that the global text prompt influences the entire image, while each local text prompt $T_l^p$ only affects the specific region defined by its corresponding local mask $M^p$.

\begin{figure*}[t]
    \centering
    \includegraphics[width=1.0\linewidth]{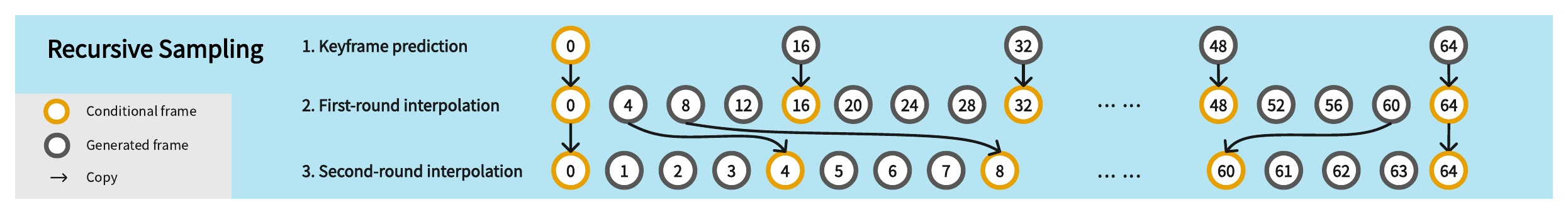}
    \caption{Recursive Sampling scheme. We sample long videos with a keyframe prediction stage followed by successive stages of frame interpolation.}
    \label{fig:recursive_sample}
\end{figure*}

\subsection{Recursive Sampling}
\label{sec:recursive_sample}
With the aforementioned Hybrid Self-Attention, our model can generate temporally consistent videos, but of fixed length $N$.
To extend the model for generating longer videos, we propose a novel Recursive Sampling scheme.
We begin by dividing a long video of length $L$ into $K$ levels.
At the $k^{\text{th}}$ level $(k=1,2,\ldots,K)$, the temporal stride between every two adjacent frames is $(N-1)^{k-1}$.
For instance, the first level $(k=1)$ consists of frames $[0, 1, 2, \ldots, L-1]$, while the second level $(k=2)$ includes frames $[0, N-1, 2(N-1), \ldots, L-1]$.
This hierarchical structure ensures that the frames at higher levels align exactly with keyframes in the adjacent lower levels.
Based on this design, we employ a two-stage inference process, where keyframe generation is first performed at the highest level $K$, followed by successive frame interpolations from level $K-1$ down to level $1$.

In \reffig{recursive_sample}, we illustrate the video sampling process with $N=5$, $K=3$, to generate a video of length $L=(N-1)^K+1=65$. 
First, at level 3, we generate keyframes $I^{16}, I^{32}, I^{48}, I^{64}$ conditioned on the reference frame $I^0$. 
Next, at level 2, we interpolate intermediate frames $I^{4}, I^{8}, I^{12}$ between $I^0$ and $I^{16}$ (and similarly in other segments). 
Finally, at level 1, we generate all remaining in-between frames to achieve full temporal resolution.
This recursive approach ensures both short-range and long-range temporal consistency across the entire sequence, enabling exponential growth in video length (e.g., merely increasing one more level to $K=4$ enables the generation of 257-frame videos).

\textbf{Adaptation to Recursive Sampling.}
This Recursive Sampling mechanism faces a key challenge: the model must learn distinct temporal correlations at each level due to varying frame intervals and differing requirements for keyframe prediction versus frame interpolation. 
A straightforward solution is to train separate models for each level, but this would lead to redundant parameters and increased training overhead.

To address this, we adapt the Hybrid Self-Attention module to handle different inputs and inference modes within a single shared model.
First, for the interpolation mode, we provide two reference latents, $r^0$ and $r^{N-1}$, and pad them with zeros in between to form the reference latent array $R$ of temporal length $N$:
\begin{equation}
    R = [r^0, \textbf{0}, \ldots, \textbf{0}, r^{N-1}]
\end{equation}
Next, we introduce switchable LoRA \cite{lora} layers for each weight in the Reference Attention: one branch for attending to the starting frame $f^0$ and another for the ending frame $f^{N-1}$ (used during interpolation). 
The Reference Attention output is then computed as:
\begin{equation}
\label{eqn:refattn_lora}
\begin{aligned}
     \text{RefAttn} \left( f^i \right)  = & \alpha_{r_0} \text{Attn} \left(  w_{r_0}^q(f^i), w_{r_0}^k(f^0), w_{r_0}^v(f^0)  \right) \\
                                        + \; \alpha_{r_1} & \text{Attn} \left(  w_{r_1}^q(f^i),  w_{r_1}^k(f^{N-1}), w_{r_1}^v(f^{N-1})  \right)
\end{aligned}
\end{equation}
where $w_{r_0}^q$ and $w_{r_1}^q$ are two separate LoRA weight attached to the original weight $w_r^q$. 
During keyframe prediction mode, we set $\alpha_{r_0} = 1.0$ and $\alpha_{r_1} = 0.0$, activating only the start-frame reference. 
During frame interpolation mode, we set $\alpha_{r_0} = \alpha_{r_1} = 0.5$, allowing the model to attend equally to both frames.
Finally, we also equip the MHF Temporal Attention with switchable LoRA layers to adapt its behavior to keyframe prediction and multiple levels in interpolation modes, ensuring appropriate temporal modeling at each recursive level.

\section{Implementation Details}

\subsection{Training Data}
Our training set comprises 235 interior scenes containing a total of 917 rooms.
For each room, we generated 120 frames by rendering along a smooth camera trajectory, resulting in 110,040 training frames.
Additionally, we incorporated the InteriorNet dataset~\cite{InteriorNet}, which provides 663 rooms with either 120 or 240 randomly selected frames per camera trajectory across 6 distinct trajectories per room, contributing a total of 730,800 frames.

\subsection{Model Training \& Inference}
During training, we input $N=5$ frames per batch and set $K=3$ levels, 
We finetune the network using the v-prediction loss function from \refeqn{loss}, with all parameters, including Hybrid Self-Attention, Masked Cross-Attention, and original layers, trainable. We train at a resolution of $640 \times 448$ using a batch size of 32 across 8 Nvidia RTX 5880 Ada GPUs for 50,000 steps, starting from a learning rate of $5.0\times10^{-5}$, decayed by 0.5 every 20,000 steps. Following XRGB \cite{xrgb}, we apply a 0.5 dropout rate for each channel. 

For inference, we perform 20-step DDIM \cite{ddim} sampling without classifier-free guidance (CFG) \cite{cfg}. The inference speed on a single Nvidia RTX 5880 Ada GPU is around 1.08 seconds per frame.
\begin{figure*}[t]
    \centering
    \includegraphics[width=1.0\linewidth]{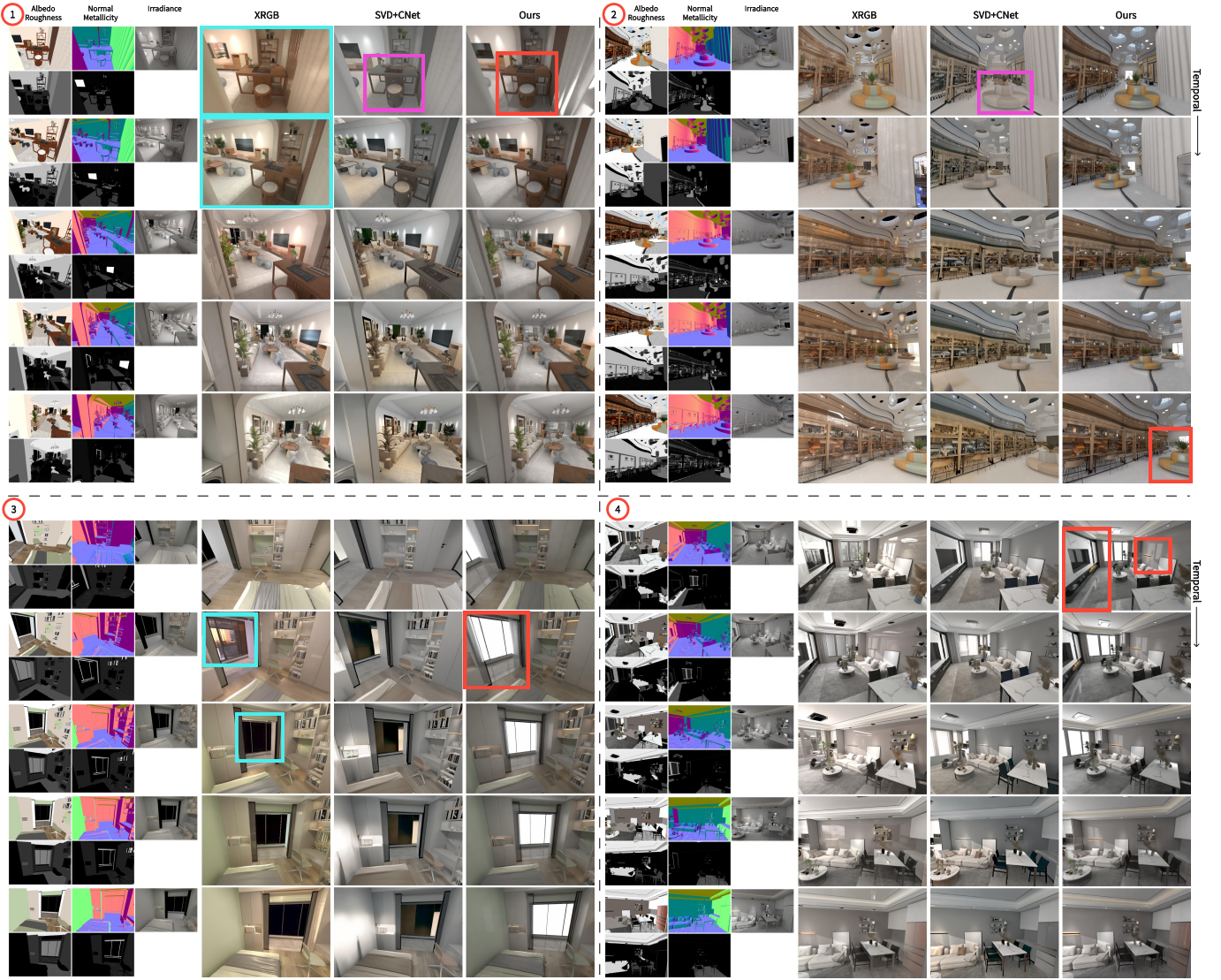}
    \caption{Qualitative comparisons, where cyan frames highlight the temporal inconsistencies observed in XRGB. Purple frames indicate wrong colors or materials produced by SVD+CNet due to the lack of intrinsic knowledge. Red frames illustrate our model's ability to infer reflective surfaces with mirrored objects. Zoom-in is recommended for better visualization.}
    \label{fig:comparison}
\end{figure*}

\begingroup
\setlength{\tabcolsep}{2.3pt}
    \begin{table}[t]
    \footnotesize
    \renewcommand{\arraystretch}{1.5}
    \centering
    \begin{tabular}{cccccccc}
         \toprule
         & FID$\downarrow$ & FVD$\downarrow$ &  PSNR$\uparrow$ &   SSIM$\uparrow$& LPIPS$\downarrow$&TC$\downarrow$ &Time$\downarrow$\\ \hline 
         XRGB&  13.3682&  672.5979&  14.6443&   0.7641& 0.1350& 3.5050&\textbf{0.89s}\\
         SVD+CNet& 23.7864& 191.9043& 16.3436& 0.8048& 0.1447& 1.0836&1.38s\\
         Ours&  \textbf{8.0562}&  \textbf{69.3828}& \textbf{22.6641}&   \textbf{0.8931}& \textbf{0.0618}& \textbf{1.0115}&1.08s\\
        PBR& -& -& -& -& -& 1.0&9.0s\\ \toprule
    \end{tabular}
    \caption{Quantitative Results on Intrinsic-Guided Models.}
    \label{tab:quantitative}
    \end{table}
\endgroup

\section{Experimental Results}

\subsection{Test Data}
Our testing set has the same structure as the training set, consisting of 237 rooms from 60 interior scenes, resulting in a total of 28,440 frames. Importantly, the scenes and rooms in the testing set do not overlap with those in the training set.

\subsection{Metrics}
As we aim for the model to generate comparable results with accurate color, material, and lighting, we use PBR-rendered videos as ground truth to evaluate the quality of the generated videos across three aspects:
\textit{1). Frame \& Video Quality:} We use \textbf{FID} \cite{fid} and \textbf{FVD} \cite{fvd} to assess the frame and video quality of the generated outputs, respectively.
\textit{2). Frame Similarity:} We measure the similarity between the generated frames and the PBR rendered frames using \textbf{PSNR}, \textbf{SSIM} \cite{ssim}, and \textbf{LPIPS}.
\textit{3). Temporal Consistency (TC):} We employ the \textbf{TC} metrics proposed in \cite{lvcd} to calculate the warping loss relative to the PBR rendered videos, which serve as the benchmark for perfect temporal consistency.

\begin{figure*}[t]
    \includegraphics[width=1.0\linewidth]{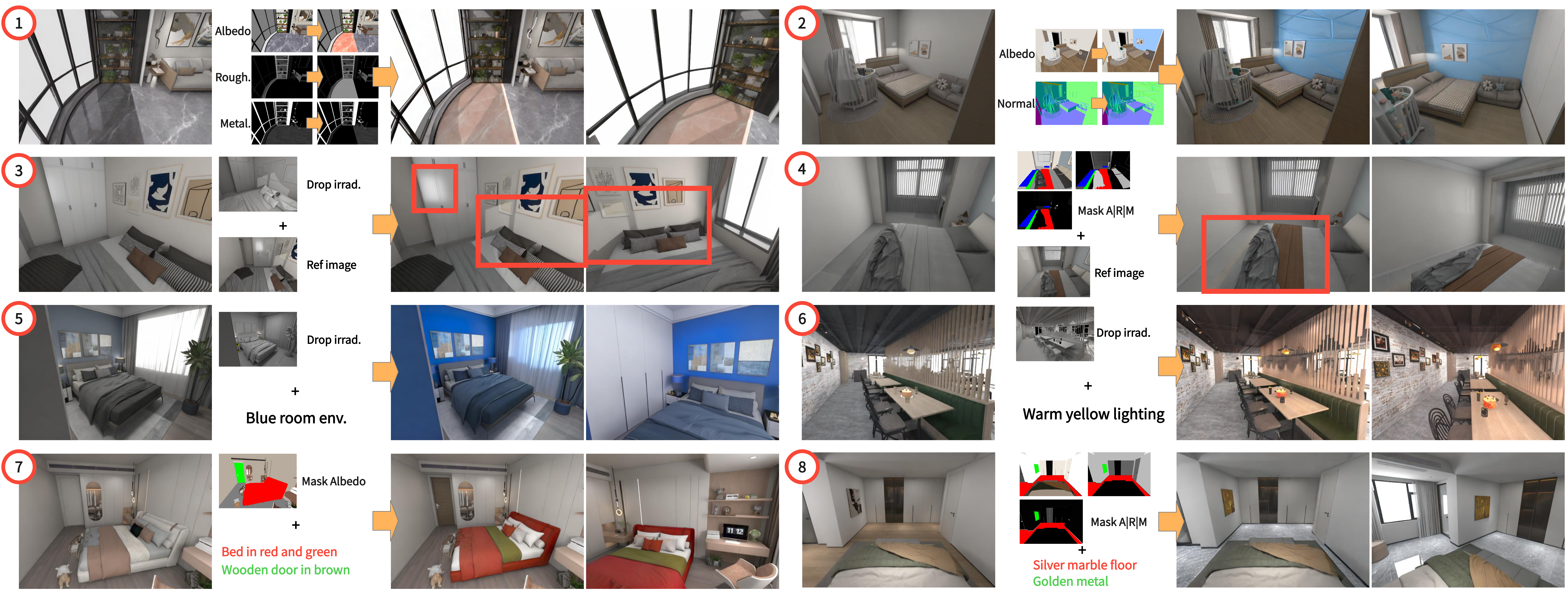}
    \caption{Qualitative results on intuitive multi-modal controls including parametric tuning on intrinsic channels (example 1 and 2), reference image control (example 3 and 4), global text control (example 5 and 6), and local text control (example 7 and 8).}
    \label{fig:multi-modal}
\end{figure*}

\subsection{Evaluations on Intrinsic-Guided Video Rendering}
We compare our method with two models. The first, XRGB \cite{xrgb}, is an intrinsic-guided image generative model. Since there is no existing video generative model that performs the same task as ours, we finetune another model (SVD+CNet) based on the pretrained Stable Video Diffusion (SVD) \cite{svd}, incorporating additional ControlNet \cite{controlnet} to accept guidance from intrinsic channels. For all three models, we feed all 5 intrinsic channels without reference or text conditions.

\textbf{Qualitative Results.}
Qualitative comparisons in \reffig{comparison} reveal XRGB's temporal inconsistency and flickering, highlighted by cyan frames in the $1^\text{st}$ and $3^\text{rd}$ examples. SVD+CNet produces over-smoothed results with inaccurate colors, shown in purple frames of the $1^\text{st}$ and $2^\text{nd}$ examples.
In contrast, our model, building upon an intrinsic-guided image model and enhanced by our Hybrid Self-Attention, generates temporally consistent videos with accurate colors.
Additionally, our model is also capable of generating reflections of light or objects based on the intrinsic knowledge, as shown in the red frames of all four examples.

\textbf{Quantitative Results.} As shown in \reftab{quantitative}, our model outperforms the other two models across all metrics.
XRGB faces challenges with temporal consistency (TC) and video quality (FVD) because it does not consider temporal correlations as an image model.
The fine-tuned video model, SVD+CNet, achieves similar temporal consistency (TC) but falls significantly short in image quality (FID) due to the lack of intrinsic knowledge without prior image pretraining.
Our model, which finetunes the pretrained image model XRGB with video data and incorporates novel Hybrid Self-Attention layers for temporal correlations, demonstrates superior performance across all metrics.

\begin{figure}[t]
    \centering
    \includegraphics[width=1.0\linewidth]{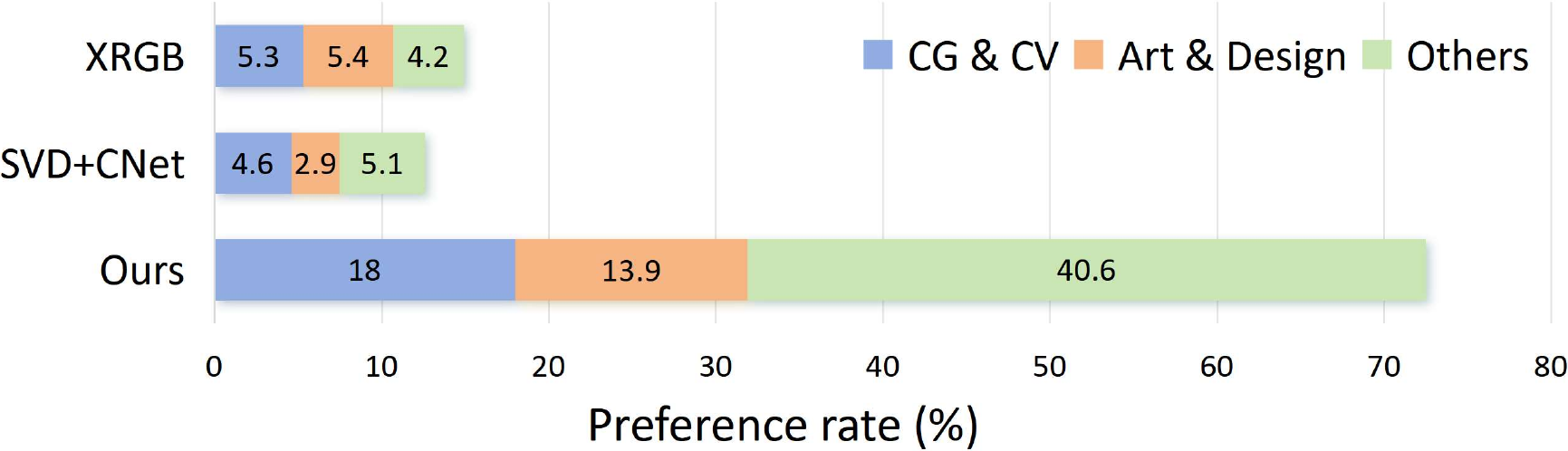}
    \caption{User study results.}
    \label{fig:user_study}
\end{figure}

\textbf{User Study.}
We further conducted a user study to evaluate the performance of our method against two other models. A pool of 20 videos was established, each containing PBR results alongside outputs generated by XRGB \cite{xrgb}, SVD+CNet \cite{svd, controlnet}, and our framework. Each participant was required to answer 14 randomly selected questions from this pool by selecting the best video results based on three aspects: 1) similarity to PBR results; 2) temporal smoothness and consistency; and 3) overall quality.
We collected 72 questionnaires from 20 participants who worked or majored in Computer Graphics and Computer Vision (CG \& CV), 16 from the Art \& Design field, and 36 from other disciplines. As shown in \reffig{user_study}, our method achieves an overall preference rate of 72.5\%.

\begingroup
\setlength{\tabcolsep}{2.6pt}
\begin{table}[t]
\footnotesize
\renewcommand{\arraystretch}{1.5}
\centering
\begin{tabular}{cccccccc}
\toprule
     & \multicolumn{3}{c}{Mask A$|$M$|$R} & & \multicolumn{3}{c}{Drop Irrad.}\\
     &   FID$\downarrow$& FVD$\downarrow$&PSNR$\uparrow$ && FID$\downarrow$& FVD$\downarrow$&PSNR$\uparrow$\\
     \midrule
      Full Intrinsics& 8.0562& 69.3828& 22.6641& & 8.0562& 69.3828&22.6641\\
    \midrule
     Mask or Drop&   10.8333&  89.1932& 20.7702&& 12.0103& 111.7795&19.0853\\
     M/D + Text&   9.9492&  80.6469& 21.6770&& 11.1065& 101.4572&20.0103\\
     M/D + Ref.&   5.9785&  49.4391& 25.3979&& 6.9414& 58.6867&23.6756\\
     \toprule
\end{tabular}
\caption{Quantitative Results on Intuitive Multi-Model Controls.}
\label{tab:multi-modal}
\end{table}
\endgroup

\subsection{Evaluations on Multi-Modal Controls}
In this section, we evaluate the effectiveness of multi-modal controls, including intrinsic channels, reference images, and text prompts.

\textbf{Qualitative Results.}
As in \reffig{multi-modal}, we first show the impact of intrinsic channels on affecting the color, material, and texture of the generated contents. In the $1^\text{st}$ example, the floor color is changed and reflections are eliminated by adjusting albedo, roughness, and metallicity.
The $2^\text{nd}$ example illustrates modifications to the texture and color of the wall.
Then in the $3^\text{rd}$ and $4^\text{th}$ examples, the reference image is used to restore the dropped or masked intrinsic information, as emphasized within the red frames.
After that, in the $5^\text{th}$ and $6^\text{th}$ examples, the irradiance channel is omitted, and text prompts are employed to indicate the overall color environment and global lighting.
Lastly, in the $7^\text{th}$ and $8^\text{th}$ examples, text prompts are used to control and modify the color and material of specific masked local regions.
The results demonstrate the effectiveness of our multi-modal conditions on controlling and editing the appearance of both global and local regions. 

\textbf{Quantitative Results.}
As shown in \reftab{multi-modal}, the results obtained using only intrinsic channels are presented as the baseline in the $1^\text{st}$ row.
Then in the $2^\text{nd}$ row, we either mask the albedo, metallic, and roughness of three random objects, or omit the irradiance channels, referring to "Mask A$|$M$|$R" and "Drop Irrad.". The observed decline across all metrics demonstrates the significance of the intrinsic channels in providing essential information for accurate outputs.
Following this, we introduce text prompts and reference images separately to compensate for the missing information caused by the masked or omitted intrinsic channels. As shown in the $3^\text{rd}$ and $4^\text{th}$ row, these additional controls lead to improvements across all metrics, which demonstrates the efficacy of our Hybrid Self-Attention and Masked Cross-Attention in utilizing reference and text conditions effectively.

\begin{figure*}[t]
    \includegraphics[width=1.0\linewidth]{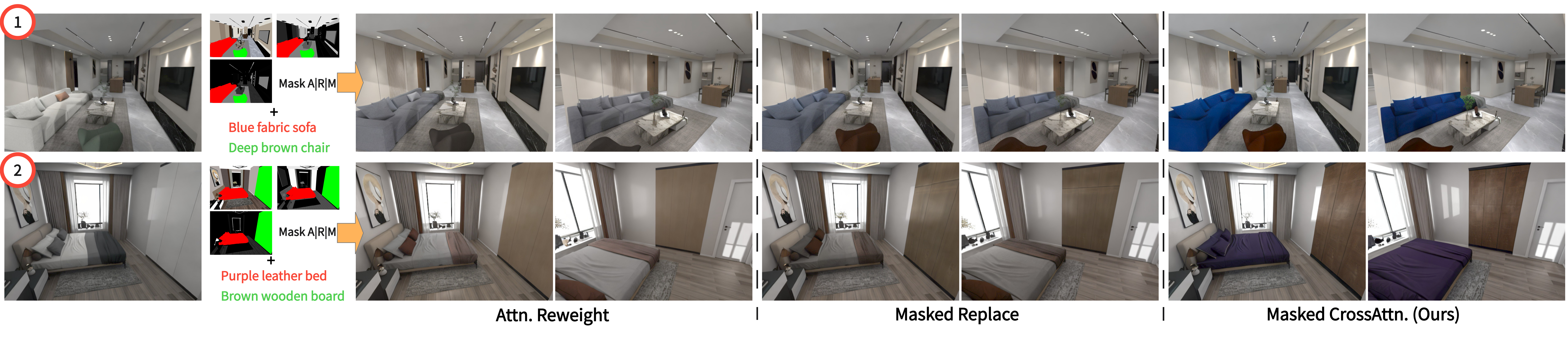}
    \caption{Ablation study on Masked Cross-Attention. Our Masked Cross-Attention can better apply the text prompts into specified local regions with more accurate colors.}
    \label{fig:abl_mask}
\end{figure*}

\subsection{Ablation Study}

\begingroup
\setlength{\tabcolsep}{5.6pt}
\begin{table}[t]
\footnotesize
\renewcommand{\arraystretch}{1.5}
\centering
\begin{tabular}{ccccc}
    \toprule
     &  FID$\downarrow$&FVD$\downarrow$&  PSNR$\uparrow$& TC$\downarrow$\\ \hline 
     1d Temp.&   8.5174&89.6895&  22.3237& 1.1105\\ 
     MHF Temp.&  \textbf{8.1460}&\textbf{79.2424}& \textbf{22.6104}&\textbf{1.0323}\\
     \midrule
     MHF Temp. + Concat&   5.7436&57.0026&  25.4121& 1.0467\\ 
     \makecell{MHF Temp. + Ref. Attn. \\ (Our Full Hybrid Attn.)}&  \textbf{5.2770}&\textbf{42.4124}& \textbf{26.2617}&\textbf{1.0165}\\
    \toprule
\end{tabular}
\caption{Ablation study on Hybrid Self-Attention.}
\label{tab:abl_temporal}
\end{table}
\endgroup

\textbf{Hybrid Self-Attention.}
In this section, we explore the impact of Multi-Head Full (MHF) Temporal Attention and Reference Attention as components of our Hybrid Self-Attention.
According to the results presented in \reftab{abl_temporal}, our MHF Temporal Attention outperforms traditional 1-dimensional temporal attention across all metrics, particularly in temporal consistency (TC) and video quality (FVD). 
Additionally, when the condition of the reference image is provided, our Hybrid Self-Attention augmented with Reference Attention outperforms basic concatenation methods in all respects.
These experiments indicate that our Hybrid Self-Attention can significantly improve both temporal consistency and fidelity to the reference image.

\begingroup
\setlength{\tabcolsep}{2pt}
\begin{table}[t]
\footnotesize
\renewcommand{\arraystretch}{1.5}
\centering
\begin{tabular}{ccccccc}
\toprule
     & \multicolumn{3}{c}{Mask A$|$M$|$R} & \multicolumn{3}{c}{Mask A$|$M$|$R + Text}\\
     &   FID$\downarrow$& FVD$\downarrow$&PSNR$\uparrow$ & FID$\downarrow$& FVD$\downarrow$&PSNR$\uparrow$\\ \hline
     Attn reweight&   10.5544&  98.0466& 20.9356& $\downarrow$ 0.4498& $\downarrow$ 6.6115&$\uparrow$ 0.2343\\
     Masked replace&   10.5441&  92.9576& 20.8315& $\downarrow$ 0.3718& $\downarrow$ 5.0223&$\uparrow$ 0.4669\\
     \makecell{Masked\\CrossAttn.(Ours)} &   10.8333&  89.1932& 20.7702& \textbf{$\downarrow$ 0.8841}& \textbf{$\downarrow$ 8.5469}& \textbf{$\uparrow$ 0.9068} \\ 
\toprule
\end{tabular}
\caption{Ablation study on Masked Cross-Attention.}
\label{tab:abl_crossattn}
\end{table}
\endgroup

\begingroup
\setlength{\tabcolsep}{3.5pt}
\begin{table}[t]
\footnotesize
\renewcommand{\arraystretch}{1.5}
\centering
\begin{tabular}{clccccccc}
\toprule
      No. of&& \multicolumn{3}{c}{Mask A$|$M$|$R} & & \multicolumn{3}{c}{Mask A$|$M$|$R + Text}\\
     Masks&& FID$\downarrow$& FVD$\downarrow$&PSNR$\uparrow$ & & FID$\downarrow$& FVD$\downarrow$&PSNR$\uparrow$\\ \hline
     0  && 8.0562 & 69.3828 & 22.6641 & & $\downarrow$0.0000 & $\downarrow$0.0000 & $\uparrow$0.0000 \\
     1  && 8.9113 & 77.8794 & 21.8109 & & $\downarrow$0.3292 & $\downarrow$5.2179 & $\uparrow$0.4809 \\
     2  && 9.9819 & 81.7982 & 21.2203 & & $\downarrow$0.6572 & $\downarrow$6.8806 & $\uparrow$0.7492 \\
     3  && 10.833 & 89.1932 & 20.7702 & & $\downarrow$0.8841 & $\downarrow$8.5469 & $\uparrow$0.9068 \\
     4  && 11.260 & 92.8582 & 20.2468 & & $\downarrow$1.1206 & $\downarrow$10.268 & $\uparrow$1.1628 \\
     5  && 11.984 & 97.5964 & 20.0262 & & \textbf{$\downarrow$1.2928} & \textbf{$\downarrow$12.896} & \textbf{$\uparrow$1.2962} \\
\toprule
\end{tabular}
\caption{Ablation study on different numbers of masks and local prompts.}
\label{tab:abl_masknumber}
\end{table}
\endgroup

\textbf{Masked Cross-Attention.}
In this section, we compare our Masked Cross-Attention with two alternatives to examine the effect of text control on local regions. 
"Attn reweight" in \cite{continuous_layout, prompt2prompt} amplifies cross-attention weights between text prompts and specific image regions.
Instead of adding the features at local regions as in \refeqn{mask_crossattn}, "Masked replace" directly replaces the features with local text attention.
As shown in \reffig{abl_mask}, qualitative results demonstrate that our approach can effectively edit the colors and materials of local regions. For instance, in the $1^{st}$ example, our method edits the colors of the sofa and chair to blue and deep brown, respectively, while the other two methods struggle to generate accurate colors. In the $2^{nd}$ example, the colors of the bed and the wardrobe are changed to purple and brown wooden by our method, while the other two methods wrongly edit the bed to brown. As for quantitative results in \reftab{abl_crossattn}, for each model, we first randomly mask the albedo, metallic, and roughness of three objects without text prompts (left group results).
We then input text prompts for each masked region and assess changes in metrics (right group results).
Our Masked Cross-Attention shows the highest improvements across all metrics.

Additionally, we conduct an ablation study to assess the stability of our method using different numbers of masks and local prompts. As shown in \reftab{abl_masknumber}, increasing the number of masks leads to a degradation in all metrics due to the loss of information (left column group). In the right column group, progressively introducing more local prompts to compensate for the missing information results in consistent improvements across all metrics, highlighting the effectiveness and stability of our local text control.

\begin{figure*}[t]
    \includegraphics[width=1.0\linewidth]{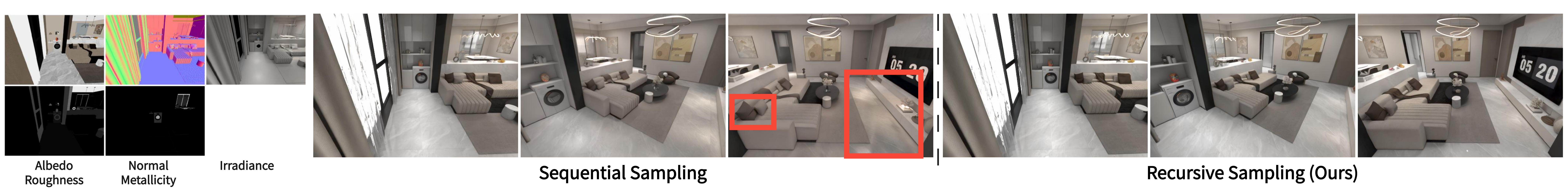}
    \caption{Ablation study on the sampling scheme. Sequential Sampling suffers from accumulated errors, where the colors of the white floor and pillows change.}
    \label{fig:abl_sample}
\end{figure*}

\begin{table}[t]
\footnotesize
\renewcommand{\arraystretch}{1.5}
\centering
\begin{tabular}{ccccc}
\toprule
     &   FID$\downarrow$& FVD$\downarrow$&PSNR$\uparrow$&TC$\downarrow$\\ \hline
     Sequential&   9.1433&  79.1601&21.5126&1.0324\\
     Recursive (Ours)&   \textbf{8.0562}&  \textbf{69.3828}&\textbf{22.6641}&\textbf{1.0115}\\ 
\toprule
\end{tabular}
\caption{Ablation study on sampling scheme.}
\label{tab:abl_sampling}
\end{table}

\textbf{Sampling Scheme.}
In this section, we compare our Recursive Sampling scheme with naive sequential sampling, in which the last frame of the previous result is used as the new reference for the next sampling. 
As shown in \reftab{abl_sampling}, our Recursive Sampling scheme outperforms sequential sampling across all evaluated aspects, including temporal consistency (TC), accuracy (PSNR), and video and frame quality (FVD and FID).
Additionally, our Recursive Sampling effectively prevents accumulated errors. As illustrated in \reffig{abl_sample}, while sequential sampling causes color changes in the floor and pillows (highlighted in red frames) due to accumulated errors, our Recursive Sampling preserves the original colors.
\section{Extensions}

\subsection{Adaptation to Other PBR Style}
As our model supports the reference image as the condition, we can use the first frame rendered by different PBRs to generate videos of different styles. As shown in \reffig{application}, while our model is trained with PBR-rendered videos featuring high brightness and low contrast (unconditional results on the left), we can employ the frame rendered with a different style as a reference to generate videos with reduced brightness and higher contrast (results on the right). 

\subsection{Generalization to Dynamic \& Outdoor Scenes}
\begin{figure*}[t]
    \centering
    \includegraphics[width=1.0\linewidth]{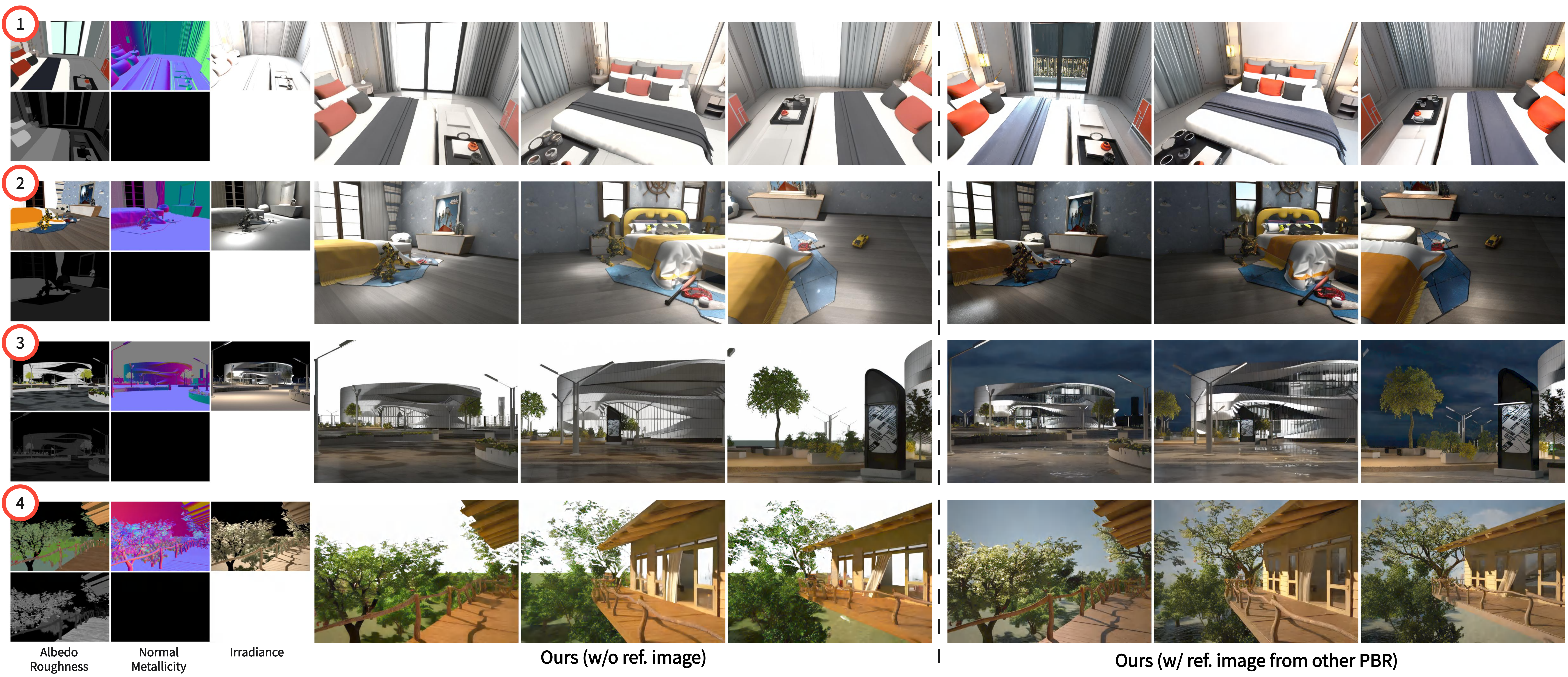}
    \caption{Extension. Our framework supports the reference frame rendered by other PBR pipelines to indicate rendering styles (examples 1-4). It can generalize to scenes with moving objects (examples 2 \& 4). With the reference frame, our model can also generalize to outdoor scenes (examples 3 \& 4).}
    \label{fig:application}
\end{figure*}

Although our model is trained exclusively on static and indoor scenes, it can generalize to both dynamic and outdoor scenes. 
As illustrated in \reffig{application}, our model can generate moving objects while animations are present in the intrinsic channels. For instance, in the $2^{\text{nd}}$ example, the robot is transforming into a car, while in the $4^{\text{th}}$ example, the curtain is moving in the wind.
Additionally, our model demonstrates generalization ability to outdoor scenes, with the exception of the sky. Since the albedo and irradiance channels for the sky are all zeros, similar to the properties of windows in our datasets, our model tends to render the sky similarly to the windows with fully white. However, this limitation can be addressed by providing a reference frame to specify the sky's appearance. 
For example, in the $3^{\text{rd}}$ and $4^{\text{th}}$ examples of \reffig{application}, the sky initially appears white without reference images but transitions to a night sky and a day sky, respectively, once reference frames are provided.

\subsection{Acceleration}
To evaluate performance and potential acceleration in practical applications where PBR-rendered frames are available as reference, we conduct experiments using PBR-rendered reference frames. As shown in \reftab{acceleration}, providing a reference frame either as the first frame or as keyframes (bypassing the first stage of keyframe generation) significantly improves all metrics, with the cost of delayed inference time. To achieve high performance while maintaining faster inference speed, we further distill a Latent Consistency Model (LCM) \cite{lcm}, enabling 2-step DDIM sampling. This allows us to achieve high performance with further acceleration.

\begingroup
\setlength{\tabcolsep}{2.5pt}
\begin{table}[t]
    \footnotesize
    \renewcommand{\arraystretch}{1.5}
    \centering
    \begin{tabular}{cccccc}
         \toprule
         & FID$\downarrow$ & FVD$\downarrow$ &  PSNR$\uparrow$ &TC$\downarrow$ &Time$\downarrow$\\ \hline
 No-Ref& 8.0562& 69.3828& 22.6641& 1.0115&1.08s\\ \hline 
         Ref-First&  5.2770&  42.4124&  26.2617& 1.0165&1.20s\\
 Ref-Keyframe& \textbf{1.9937}& \textbf{29.5864}& \textbf{31.8192}& \textbf{1.0157}&1.69s\\ \hline
         Ref-First + LCM 2-step&  5.4832&  45.6137&  25.1470& 1.0297&\textbf{0.24s}\\
 Ref-Keyframe + LCM 2-step& 2.1381& 31.0491& 31.1551& 1.0289&0.79s\\
    \toprule
    \end{tabular}
    \caption{Acceleration with PBR-rendered reference.}
    \label{tab:acceleration}
\end{table}
\endgroup

\begin{figure*}[t]
    \centering
    \includegraphics[width=1.0\linewidth]{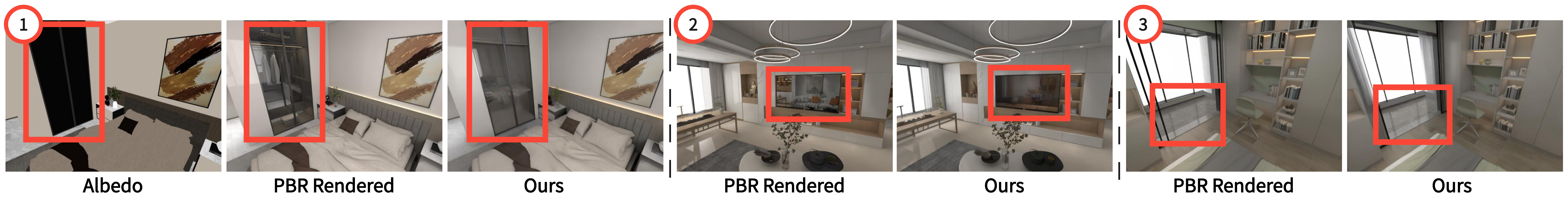}
    \caption{Limitations. Our method struggles to generate: 1) contents behind transparent glass which are not indicated by intrinsic channels (example 1); 2) sharp contents in highly reflective surfaces (example 2).}
    \label{fig:limitation}
\end{figure*}

\section{Conclusion \& Limitations}
In summary, we present the first intrinsic-guided video generation framework that supports multi-modal controls, enabling flexible and precise video rendering guided by reference images and text prompts applied to both global and local regions.
Building upon the XRGB, we introduce a novel Hybrid Self-Attention mechanism that improves temporal coherence and enhances fidelity to reference frames. 
Furthermore, we propose a novel Masked Cross-Attention that allows textual control over global and specified local regions by selectively attending to relevant spatial regions.
To enable the generation of long videos with temporal consistency, we develop a new Recursive Sampling strategy that incorporates progressive frame sampling, effectively reducing error accumulation.
Finally, to enable training of intrinsic-guided video generation, we introduce the InteriorVideo dataset, featuring interior scenes with smooth camera trajectories and reliable intrinsic channels.
Extensive experiments demonstrate that our framework can generate long, temporally consistent, and photorealistic videos with precise material, geometry, and lighting based on intrinsic conditions.
Moreover, our approach supports multi-modal controls, including reference images, global and local text prompts, as well as video editing through parametric tuning, exploring new possibilities for video rendering through intuitive video synthesis.

Despite the high quality of our framework, there are certain limitations.
As shown in the $1^{\text{st}}$ example of \reffig{limitation}, the model struggles to synthesize content behind transparent glass surfaces, as the input intrinsic channels only capture the properties of the glass itself. 
To mitigate this, future work could incorporate explicit modeling of transparency during intrinsic channel estimation, enabling more accurate reconstruction of occluded regions.
Additionally, while our model can render plausible reflections for light sources or objects spatially close to reflective surfaces (as demonstrated in the $3^{\text{rd}}$ example), it fails to generate sharp reflections of distant or out-of-view objects, as seen in the $2^{\text{nd}}$ example.
This limitation highlights a fundamental challenge in 2D intrinsic-guided generation: the lack of complete 3D scene understanding.
Addressing this may require integrating explicit 3D scene representations into the generation process to better synthesize realistic reflections.

\bibliographystyle{IEEEtran}
\bibliography{bibliography}

\end{document}